\documentclass[11pt]{article}
\usepackage{amsmath,amssymb,amsthm}
\usepackage{geometry}
\usepackage{booktabs}
\usepackage{hyperref}
\hypersetup{hidelinks} 

\title{A Contextual Seven-Valued Logic (\emph{Saptabhaṅgīnaya}) for Quantum Systems}
\author{Partha Ghose \footnote{partha.ghose@gmail.com}\\Tagore Centre for Natural Sciences and Philosophy,\\ Rabindra Tirtha, New Town, Kolkata 700156, India}
\date{}

\begin{document}
\maketitle
\begin{abstract}
The quantum measurement problem is often presented as a conflict between unitary evolution and non-unitary collapse. Drawing on Wittgenstein’s later philosophy of language and Bohr’s principle of complementarity, we argue that this conflict is a grammatical illusion arising from cross-context conflations. To address this, we introduce a contextual seven-valued logic modeled on the Jaina doctrine of \emph{saptabhaṅgīnaya} (sevenfold predication). In one formulation, each proposition is assigned a triplet $(t,f,u)$ indicating its status as true, false, or unsayable within a given context, with paraconsistent rules blocking triviality. In another, contexts are explicitly formalized through quantified conditionals, aligning directly with Bohr’s view that meaning derives from experimental arrangements. By comparing these two complementary approaches, we show how canonical paradoxes—including Schrödinger’s cat and Wigner’s friend—dissolve once context is made explicit. The result is a flexible logical framework that reconciles Wittgensteinian conceptual therapy, Bohr’s complementarity, and the Jaina pluralistic tradition, offering a coherent semantics for quantum discourse.
\end{abstract}
\section{Introduction}
Wittgenstein's later philosophy, especially in the \emph{Philosophical Investigations}, rejects the earlier ``picture theory'' of meaning from the \emph{Tractatus} and replaces it with the dictum that \emph{meaning is use}. Words acquire significance inside \emph{language-games}, practices embedded in \emph{forms of life}. Philosophical confusion arises when we detach expressions from their proper games and attempt to deploy them on what he calls the ``frictionless ice'' of abstraction, where language no longer grips. The therapeutic task is to return to the ``rough ground'' of ordinary use, recovering the grammatical rules that make our concepts work.

Bohr's principle of complementarity can be read in this Wittgensteinian light. ``Wave'' and ``particle'' are not rival metaphysical descriptions of a single essence; they are context-bound grammars tied to experimental arrangements. Questions like ``Is the electron \emph{really} a wave or a particle?'' misapply grammar across games. The measurement problem---the apparent clash between unitary evolution and collapse---likewise emerges from sliding between distinct uses. The point is not to eliminate physics, but to mark that our \emph{descriptions} are context-sensitive, and that forcing a single global grammar manufactures paradox.

The aim of this note is to turn this therapeutic insight into a succinct \emph{calculus}: a seven-valued, explicitly context-indexed paraconsistent logic reflecting the classical Jaina saptabhaṅgī predications. This logic prevents cross-context explosions and provides a disciplined way to speak about quantum scenarios (double-slit, Schr\"odinger’s cat, Wigner’s friend) without demanding a single, context-free valuation.

\section{Seven Context-Dependent Truth-Values}
In the Jaina system, every assertion is qualified by ``syāt'' (``in some respect / context''). We formalize this using non-empty triples $(t,f,u) \in \{0,1\}^3 \setminus \{(0,0,0)\}$, read relative to context $c$ as:
\begin{itemize}
  \item $t=1$ : syāt \emph{asti} (true-in-$c$)
  \item $f=1$ : syāt \emph{nāsti} (false-in-$c$)
  \item $u=1$ : syāt \emph{avaktavyam} (indescribable-in-$c$)
\end{itemize}

This yields the seven saptabhaṅgī values:
\begin{align*}
v_1 &= (1,0,0) & \text{syāt asti} \\
v_2 &= (0,1,0) & \text{syāt nāsti} \\
v_3 &= (1,1,0) & \text{syāt asti-nāsti} \\
v_4 &= (0,0,1) & \text{syāt avaktavyam} \\
v_5 &= (1,0,1) & \text{syāt asti-avaktavyam} \\
v_6 &= (0,1,1) & \text{syāt nāsti-avaktavyam} \\
v_7 &= (1,1,1) & \text{syāt asti-nāsti-avaktavyam}
\end{align*}

\section{Connectives}
We define connectives componentwise on $(t,f,u)$.

\subsection*{Negation}
\begin{equation}
\neg(t,f,u) = (f,t,u).
\end{equation}

\subsection*{Conjunction}
\begin{equation}
(t,f,u) \wedge (t',f',u') = (t \wedge t', \; f \vee f', \; u \vee u' \vee (t \wedge u') \vee (u \wedge t')).
\end{equation}

\subsection*{Disjunction}
\begin{equation}
(t,f,u) \vee (t',f',u') = (t \vee t', \; f \wedge f', \; u \vee u' \vee (f \wedge u') \vee (u \wedge f')).
\end{equation}

\subsection*{Implication}
Defined via material reduction:
\begin{equation}
P \to Q := \neg P \vee Q.
\end{equation}

\section{Designatedness and Entailment}
A value is \emph{designated} in context $c$ iff $t=1$. For $\Gamma, \Delta$ sets of formulas,
\begin{equation}
\Gamma \Rightarrow \Delta [c]
\end{equation}
is valid iff for every valuation $V(\cdot,c)$, either some $A \in \Gamma$ is non-designated ($t=0$), or some $B \in \Delta$ is designated ($t=1$). This yields a paraconsistent calculus: from $P$ and $\neg P$ one cannot infer arbitrary $Q$.

\section{Dynamics Within and Between Contexts}
Within a fixed context $c$, states evolve under $U_t^c$, corresponding to deterministic dynamics. Measurement outcomes are modeled as context-switch operators
\begin{equation}
M_\alpha : c \to c'.
\end{equation}
Thus, unitary evolution and ``collapse'' are not contradictory laws but distinct operators: $U_t^c$ acts within $c$, while $M_\alpha$ changes the context index.

\section{Worked Examples}
\subsection*{Double-Slit}
Let $P_A$ = ``particle went through slit A'' and $I$ = ``interference observed.''
\begin{itemize}
  \item In $c_{\text{wp}}$ (which-path): $V(P_A,c_{\text{wp}}) = (1,0,0)$, $V(I,c_{\text{wp}}) = (0,1,0)$.
  \item In $c_{\text{wave}}$ (interference): $V(I,c_{\text{wave}}) = (1,0,0)$, $V(P_A,c_{\text{wave}}) = (0,0,1)$.
\end{itemize}
Thus, asking ``Which slit?'' in the interference context yields \emph{syāt avaktavyam}, not a contradiction.

\subsection*{Spin-$\tfrac{1}{2}$ in Incompatible Bases}
Let $S_z$ = ``spin up in $z$ basis'' and $S_x$ = ``spin up in $x$ basis.''
\begin{itemize}
  \item In $c_z$: $V(S_z,c_z) = (1,0,0)$, $V(S_x,c_z) = (0,0,1)$.
  \item In $c_x$: $V(S_x,c_x) = (1,0,0)$, $V(S_z,c_x) = (0,0,1)$.
\end{itemize}
Thus, incompatibility of observables is expressed as ``unsayability'' in the wrong context.

\section{Schr\"odinger’s Cat and \emph{Saptabhaṅgīnaya}}
The cat thought experiment dramatizes the tension between quantum superposition and classical definiteness. Before observation, the composite (nucleus+detector+cat) is modeled as a superposed state correlating ``decayed/triggered/dead'' and ``not-decayed/not-triggered/alive'' branches. In classical logic, saying ``the cat is alive and dead'' is absurd. In the \emph{saptabhaṅgīnaya} semantics, the proposition \emph{Alive} receives a contextual value: relative to the pre-observation context $c_{\text{closed}}$, $V(\text{Alive},c_{\text{closed}})$ may be $(1,1,0)$ (sy\={a}t asti--n\={a}sti) or $(0,0,1)$ (sy\={a}t avaktavyam), capturing that the ordinary predicate does not apply as it does post-observation. Opening the box effects a context-switch $M_\alpha:c_{\text{closed}}\!\to\! c_{\text{open}}$; in $c_{\text{open}}$ one obtains a classical predication $V(\text{Alive},c_{\text{open}})=(1,0,0)$ or $(0,1,0)$. The ``paradox'' is thus a cross-context misuse, not a logical inconsistency.

\section{Wigner’s Friend and \emph{Saptabhaṅgīnaya}}
Wigner’s friend measures inside a sealed lab and records a definite outcome. For the friend’s context $c_{\text{friend}}$, the proposition ``Outcome = $o$'' is designated: $V(\text{Outcome}=o, c_{\text{friend}})=(1,0,0)$. For Wigner, who treats the entire lab as a quantum system in context $c_{\text{Wigner}}$, the corresponding proposition carries a superposed/entangled reading, e.g. $(1,1,0)$ or $(0,0,1)$. Classical logic forces a clash: both cannott be right. Context-indexed \emph{saptabhaṅgīnaya} removes the demand for a single global valuation. Each report is correct \emph{in its context}; bridge inferences across contexts require explicit rules (the context-switch operator $M_\alpha$ or a compatibility map), preventing illicit globalization of one description.

\section{Comparison with Quantificational Context Logic}

In addition to the triplet semantics developed above, an alternative formulation of
\emph{saptabhaṅgīnaya} logic has been proposed using explicit quantification over contexts (Ghose and Patra 2023, 2025). 

Using the quantifier $\forall$ the three basic conditional truth values can be written as

(i) $\forall x\, [\phi(x) \rightarrow p(x)]$; 

(ii) $\forall x\, [\phi(x) \rightarrow \neg p(x)]$; 

(iii) $\forall x\, [\phi(x) \rightarrow q(x)]$.

Here, $x$ is a variable ranging over a domain of discourse (e.g., clay pots), $\phi$ is a well-formed formula specifying a context or condition (such as `is baked'), $p$ is a predicate (e.g., `is red'), and $q$ is the predicate expressing \emph{avaktavyam}.

To illustrate: the first formula, $\forall x\, [\phi(x) \rightarrow p(x)]$ means, `For all $x$ (say, clay pots), if $x$ satisfies the condition $\phi$ (e.g., is baked), then $x$ is red'. The logic retains explicit reference to the condition under which the truth of a statement is evaluated.

The other four compounds can be written as 

(iv) $\forall x\,[\phi(x) \rightarrow p(x) \land \phi^\prime(x)\rightarrow \neg p(x)] \land \neg [ \phi(x) \leftrightarrow \phi^\prime (x)]$,  

(v) $\forall x\,[\phi(x) \rightarrow p(x) \land \phi^\prime(x)\rightarrow q(x)] \land \neg [ \phi(x) \leftrightarrow \phi^\prime (x)]$,  

(vi) $\forall x\,[\phi(x) \rightarrow \neg p(x) \land \phi^\prime(x)\rightarrow q(x)] \land \neg [ \phi(x) \leftrightarrow \phi^\prime (x)]$,  
 
(vii) $\forall x\,[\phi(x) \rightarrow p(x) \land \phi^\prime(x)\rightarrow \neg p(x) \land \phi^{\prime\prime}(x) \rightarrow q(x)] \land \neg [\phi(x) \leftrightarrow \phi^\prime (x)] \land\neg[ \phi^\prime (x) \leftrightarrow \phi^{\prime\prime}(x)] \land \neg[\phi(x)\leftrightarrow \phi^{\prime\prime}(x)]$.

When expressed in this formal framework, the seven predications are mutually consistent, as each holds under distinct, non-overlapping conditions.

Let us briefly compare the two.

In the Triplet Semantics (Algebraic/Paraconsistent) Logic,
each formula $P$ in a context $c$ is evaluated to a triple $(t,f,u)$ with $t=1$ meaning
true-in-$c$, $f=1$ false-in-$c$, and $u=1$ unsayable-in-$c$. The seven values are the
non-empty subsets of $\{t,f,u\}$. Logical connectives are defined componentwise, and
designatedness is $t=1$, yielding a paraconsistent calculus that blocks explosion.

In the Quantificational Context Logic (Bohr-style), truth is expressed with quantified conditionals whose antecedents explicitly denote contexts. With domain variable $x$, context formula $\varphi(x)$, predicate $p(x)$ and
indescribability predicate $q(x)$, the three basic values are:
\begin{align*}
(T)&:\ \forall x [\varphi(x)\to p(x)],\\
(F)&:\ \forall x [\varphi(x)\to \neg p(x)],\\
(U)&:\ \forall x [\varphi(x)\to q(x)].
\end{align*}
The other four arise by conjoining clauses under mutually incompatible conditions
$\varphi,\varphi',\varphi''$. Contradictions are avoided because different values are tied to
distinct contexts.

The conceptual differences can thus be summarized as:
\begin{enumerate}
\item In the triplet system, context is an index on valuations; in the quantificational system, context
is syntactic in the antecedent.
\item The triplet system allows values like $(1,1,0)$ within a single context, controlled by paraconsistency.
Quantificational logic distributes $T$ and $F$ across incompatible contexts, avoiding direct
contradiction.
\item The triplet system resembles algebraic paraconsistent logics; quantificational logic mirrors Bohr’s Complementarity
Principle more directly by writing the experimental arrangement into the syntax.
\end{enumerate}

\subsection*{Summary Table}
\begin{center}
\begin{tabular}{@{}p{0.3\linewidth}p{0.3\linewidth}p{0.3\linewidth}@{}}
\toprule
 & \textbf{Triplet semantics} \& \textbf{Quantificational logic} \\
\midrule
Truth-bearers & Triples $(t,f,u)$ & $\forall x[\varphi\to\cdot]$ clauses \\
Seven values & Subsets of $\{t,f,u\}$ & Conjunctions of distinct $\varphi,\varphi',\varphi''$ \\
Connectives & Componentwise & Classical, but context-restricted \\
Contradiction & Direct, paraconsistently tamed & Distributed across contexts \\
Bohr alignment & Implicit via indexing & Explicit via conditions \\
\bottomrule
\end{tabular}
\end{center}

The triplet calculus is compact and algebraic, suitable for automated or algebraic reasoning.
The quantificational version is closer to Bohr’s original intent, useful for expository clarity
and protocol-level analyses of context. Together they provide complementary perspectives.
\section{Conclusion}
The two formulations of \emph{saptabhaṅgīnaya} logic—the algebraic triplet semantics and the quantificational context logic—should be seen not as rivals but as complementary tools. The triplet approach provides a compact algebraic calculus suitable for paraconsistent reasoning and formal proof systems, while the quantificational approach makes Bohr’s insight about the primacy of experimental arrangements transparent at the syntactic level. Together, they embody the pluralistic spirit of the Jaina doctrine while giving it precise logical shape. By integrating these perspectives, we obtain a richer account of contextual truth that avoids paradox without sacrificing descriptive power, and which may be extended beyond quantum mechanics to other domains where context and apparent contradiction are inseparable.
\section{Acknowledgements}
I am grateful to the Director of the Indian Institute of Technology (IIT) Mandi for providing a stimulating intellectual environment amidst the magnificent Himalayas, where this paper was conceived and written.

\end{document}